\documentclass[traditabstract,printer]{aa}
\usepackage{txfonts}
\usepackage{graphicx}
\usepackage{subfigure}
\usepackage{color}
\usepackage{natbib}
\newcommand\phs{\phantom{$-$}}

\newcommand{\rb}[1]{\raisebox{1.5ex}[-1.5ex]{#1}}
\begin{document}
\title{Kinematics of outer halo globular clusters: M\,75 and  NGC\,6426\thanks{Based on observations collected at the European Southern Observatory under 
proposals 69.B-0305 (P.I. E. Tolstoy) and proposal 99.B-0012 (P.I. A. Koch).   This paper includes data gathered with the 6.5 meter Magellan/Clay Telescope located at Las Campanas Observatory, Chile.
}
} 

\author{
Andreas Koch\inst{1} 
  \and Michael Hanke\inst{1}
  \and Nikolay Kacharov\inst{2}
  }
  
\authorrunning{A. Koch et al.}
\titlerunning{Kinematics of outer halo clusters: M\,75 and NGC\,6426}
\offprints{A. Koch;  \email{andreas.koch@uni-heidelberg.de}}
\institute{Zentrum f\"ur Astronomie der Universit\"at Heidelberg, Astronomisches Rechen-Institut, M\"onchhofstr. 12, 69120 Heidelberg, Germany
\and Max-Planck-Institut f\"ur Astronomie, K\"onigstuhl 17, 69117 Heidelberg, Germany
}
\date{}
\abstract{
Globular clusters (GCs) and their dynamic interactions with the Galactic components provide an important insight into the structure and  
formation of the early Milky Way. 
Here, we present a kinematic study of two outer halo GCs based on a combination of VLT/FORS2, VLT/FLAMES, and Magellan/MIKE low- and high-resolution spectroscopy 
of 32 and 27 member stars, respectively. 
Although both clusters are located at Galactocentric distances of 15 kpc, they have otherwise very 
different properties. 
M\,75 is a luminous and metal-rich system at [Fe/H] = $-1.2$ dex, a value that we confirm from the calcium triplet region. 
This GC shows { mild} evidence for rotation with an amplitude of  $A_{\rm rot}\sim$5 km\,s$^{-1}$. 
One of the most metal-poor GCs in the Milky Way (at [Fe\,{\sc ii}/H] = $-2.3$ dex), NGC\,6426 exhibits { marginal} evidence of internal rotation at the 
2 km\,s$^{-1}$ level. Both objects have velocity dispersions that are consistent with their luminosity. 
{ Although limited by small-number statistics}, the resulting limits on their $A_{\rm rot}/\sigma_0$ ratios suggest that M\,75 is
a slow rotator driven by internal dynamics rather than being effected by the weak Galactic tides at its large distances. 
Here, M\,75 ($A_{\rm rot}/\sigma=0.31$) is fully consistent with the properties of 
other, younger halo clusters. 
At $A_{\rm rot}/\sigma_0=0.8\pm0.4$, NGC\,6426 appears to have a remarkably ordered internal motion for its low metallicity, but { the large uncertainty
does not allow for an unambiguous categorization as a fast rotator.} 
An accretion origin of M\,75 cannot be excluded, { based on the eccentric orbit, which we derived from the recent data release 2 of Gaia, 
and considering its younger age.} 
}
\keywords{Globular clusters: individual: M\,75, NGC\,6426 --- Galaxy: halo --- Galaxy: kinematics and dynamics --- Galaxy: structure}
\maketitle 
%
%
%
%
\section{Introduction}
Being amongst the oldest stellar systems in the universe, globular clusters (GCs) are ideal tracers of the earliest phases of
the Milky Way's assembly. A particular emphasis lies on objects at large Galactocentric distances, as they probe 
the outermost halo and its accreted, ex-situ origin \citep{SearleZinn1978,Hartwick1987,Carollo2007,Pillepich2015}, or the 
important transition region between the inner and outer Galactic halo.
Despite their, on average, younger ages \citep{Mackey2004,Marin-Franch2009}, outer halo GCs are chemically very similar to halo clusters at smaller 
Galactocentric radii \citep{Koch2009,Koch2010,Caliskan2012}. 

{  Internally, GCs are ever more deviating from the historic view of simple stellar populations. 
Similarly, spherical symmetry and dynamic equilibrium are often too simplistic assumptions for many of these objects}. 
It is nowadays well established that these object host multiple populations, indicating a great degree of complexity during their early formation epochs 
\citep{Decressin2007,DErcole2008,Bastian2013,Milone2017,Martocchia2018},  as
is presently most pronounced in terms of their light chemical abundance variations \citep{Osborn1971,Kayser2008,Carretta2009NaO,Gratton2012}. 
However, none of the models presents a fully satisfactory formation picture, yet \citep{Bastian2015,Bastian2017}.

In this context, a spatial segregation between first and second generations of stars has been observed \citep[e.g.,][]{Carretta20103201}, which is often 
accompanied by differences in their internal kinematics such as significantly different velocity dispersions and rotational amplitudes \citep{Bellazzini2012}. 
Furthermore, systematic correlations of the kinematic properties with other GC-characteristics, such as their metallicity or age have been reported 
\citep{Carretta2010,Bellazzini2012,Kacharov2014,Lardo2015}.
All these can be considered an imprint of the GCs' formation history rather than being a result of  long-term dynamical evolution, so that 
detailed dynamical models are in place \citep{Bianchini2013,Vesperini2013,Cordero2017,Baumgardt2017}. 
Rotation has by now been observed in most GCs and it leads to their morphological flattening, although also external effects such as pressure anisotropy and the 
{ Galactic tidal field can play a significant role in affecting the 
dynamical evolution and shape of these systems \citep[e.g.,][]{Tiongco2018}.} 
Observationally, systematic monitoring of the internal dynamics of GCs has started blooming owing to the large multiplexing capacities of multi-object spectrographs and
with the latest developments in integral-field unit spectroscopy  \citep[e.g.,][]{Kamann2018,Ferraro2018}. 

In this paper, we determine the kinematic properties of two GCs that do not have many 
common characteristics (such as metallicity or mass) except for their location in the transition region between the inner and outer Galactic halos, 
at R$_{\rm GC}$=14.4 and 14.6 kpc, respectively. 
M\,75 ($\equiv$NGC\,6864) is a massive, metal-rich ([Fe/H]=$-1.2$ dex) GC that exhibits an extraordinary, trimodal 
horizontal branch (HB) morphology \citep{Catelan2002,Kacharov2013}. In contrast, the chemical abundance study of  \citet{Kacharov2013} 
revealed only moderate light element anti-correlations, lacking a third, ``extreme'' component that would go in lockstep with the 
extended, hot HB, concluding that this object's HB peculiarity has been shaped by processes that are yet unknown.  
On the other hand, the low-mass NGC\,6426 is one of the three most metal-poor and oldest GCs in the Milky Way system \citep{Salaris2002,Hanke2017}. 
Owing to a limited spectral sample, \citet{Hanke2017} could not establish the presence of a Na-O anti-correlation in NGC\,6426, but mild variations 
in other elements suggested that multiple populations are  also likely present in this object. 
Both our target clusters are, kinematically, hitherto uncharted.

This paper is organized as follows:
Sect.~2 presents the data and our velocity measurements. Additional measurements of metallicities for the M\,75 stars are briefly discussed in Sect.~3,
before describing the general kinematic properties of the two GCs  in Sect.~4. Next,  Sect.~5 focuses on 
an in-depth analysis of the clusters' internal rotation,  before discussing our 
findings in Sect.~6. 
\section{Data and radial velocity measurements}
\subsection{M\,75: MIKE high-resolution spectra}
The spectra used in the present analysis are those already  presented in \citet{Kacharov2013}. In brief, high-resolution (R$\sim$40,000) spectra 
of 18 red giant member stars have been obtained with the Magellan Inamori Kyocera Echelle (MIKE) spectrograph at the 6.5-m Magellan2/Clay Telescope 
at Las Campanas Observatory, Chile.
Out of these, 16 had a sufficient signal-to-noise ratio (SNR) for a detailed abundance analysis. 
We measured their radial velocities via a cross-correlation of the spectra against a synthetic template of a red giant branch (RGB) star of stellar parameters typical of the observed sample. 
\subsection{M\,75: VLT/FORS2 low-resolution data}
Low-resolution spectroscopy of 37 stars within 3.8$\arcmin$ (7.5 half-light radii $r_h$) of the GC is available from the ESO archive. 
The location of these targets in a colour magnitude diagram (CMD) is shown in Fig.~1. 
\begin{figure}
\centering
\includegraphics[width=0.49\hsize]{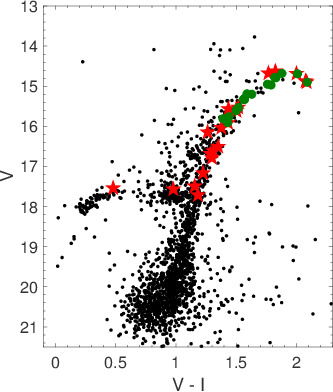}
\includegraphics[width=0.49\hsize]{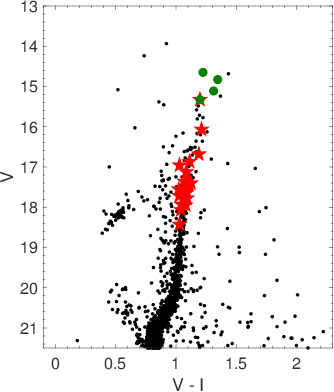}
\caption{Left: CMD of M75 from \citet{Kravtsov2007} with spectroscopic targets shown as red star symbols (FORS members), 
and green circles (MIKE targets). Right: CMD of NGC\,6426 from the HST photometry of \citet{Dotter2011}, indicating member candidates (red stars). Here, 
green circles show the four stars studied in high 
resolution by \citet{Hanke2017}.}
\end{figure}
These spectroscopic data have been taken during one night in Aug 2002 with the Focal Reducer and low dispersion Spectrograph FORS2 at the Very Large Telescope (VLT) under proposal 69.B-0305 (P.I. E. Tolstoy). The filter used was OG590 with grism 1028z, centered around the near-infrared calcium triplet (CaT) lines at $\sim$8500 \AA, and the exposure time was 2$\times$900 s. 
These data were reduced using IRAF's {\em apall} and other standard routines \citep[see also][]{Fraternali2009} and have a resolving power of $\sim$2500, as determined from the width of the sky emission lines. Typical SNR values range from 7--120 pixel$^{-1}$, with a median  of $\sim$70 
pixel$^{-1}$.

Heliocentric velocities of the target stars were determined by cross correlation (within IRAF's {\em fxcor}) against a synthetic spectrum of the three CaT lines, adopted as simple Gaussian profiles \citep{Kleyna2004}. One of the stars is a blue HB star and we measured its velocity from the Doppler shifts of the prominent, broad Paschen lines, yielding a larger uncertainty. 
The formal errors returned by {\em fxcor} are of the order of 1.5 km\,s$^{-1}$. 
As a result (Fig.~2), we find 21  stars within 3.3$\arcmin$ (6.7 r$_h$) 
with velocities below $-150$ km\,s$^{-1}$ that are clearly separated from the bulk of Galactic foreground stars. 
We therefore assume these as GC member candidates.  Their properties are listed in 
Table~1 and Fig.~1 shows their spatial distribution and velocity histogram. 
\begin{figure}
\centering
\includegraphics[width=1\hsize]{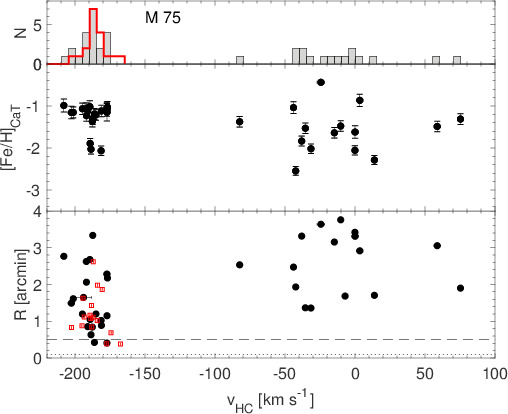}
\caption{Heliocentric radial velocity against radial distance (bottom panel) and CaT-based metallicity (middle) from the FORS data are shown in black. 
The histogram in the top panel indicates the peak of M\,75 at $-$188 km\,s$^{-1}$,  
and the dotted and dashed lines in the bottom panel  show the cluster's core- and half-light radii. 
Red symbols in the bottom panel and the red solid line in the histogram plot refer to the MIKE sample.}
\end{figure}
\begin{table*}
\caption{Properties of M\,75 member stars.}
\centering          
\renewcommand{\footnoterule}{}  
\begin{tabular}{rccccccrccr}
\hline
\hline
 & $\alpha$ & $\delta$ & B & V & I & \multicolumn{2}{c}{v$_{\rm HC}$ [km\,s$^{-1}$]} & $\Sigma$W & [Fe/H]$_{\rm CaT}$ & \\
\raisebox{1.5ex}[-1.5ex]{ID$^a$} & \multicolumn{2}{c}{(J2000.0)} & [mag] & [mag] & [mag] & FORS2 & MIKE & [m\AA] & [dex] 
& \raisebox{1.5ex}[-1.5ex]{Type}\\
\hline
\multicolumn{10}{c}{FORS2}\\
\hline
 652 & 20 05 56.35 & $-$21 54 35.49 & 16.42 & 14.63 & 12.80 &  $-$191.8$\pm$1.0 &            \dots & 5.75$\pm$0.10 & $-$1.08$\pm$0.12 &	RGB \\
 583 & 20 06 07.96 & $-$21 54 52.81 & 16.34 & 14.68 & 12.91 &  $-$187.6$\pm$0.6 & $-$188.1$\pm$0.1 & 5.43$\pm$0.11 & $-$1.20$\pm$0.12 &	RGB \\
 528 & 20 06 00.24 & $-$21 55 12.57 & 16.70 & 14.69 & 12.68 &  $-$189.2$\pm$0.7 & $-$188.7$\pm$0.2 & 4.33$\pm$0.09 & $-$1.66$\pm$0.10 &	RGB \\
 461 & 20 06 03.80 & $-$21 55 34.60 & 16.62 & 14.78 & 12.96 &             \dots & $-$167.5$\pm$0.1 & 	     \dots &	        \dots & RGB \\
 503 & 20 06 10.89 & $-$21 55 23.10 & 16.69 & 14.79 & 12.96 &             \dots & $-$188.3$\pm$0.1 & 	     \dots &	        \dots & RGB \\
 391 & 20 06 05.47 & $-$21 55 52.82 & 16.82 & 14.87 & 12.79 &  $-$188.6$\pm$0.7 &            \dots & 3.92$\pm$0.09 & $-$1.78$\pm$0.10 &	RGB \\
1307 & 20 06 02.83 & $-$21 54 20.98 & 16.93 & 14.90 & 12.81 &  $-$181.3$\pm$1.7 & $-$184.1$\pm$0.2 & 3.83$\pm$0.08 & $-$1.81$\pm$0.10 &	RGB \\
1312 & 20 06 06.79 & $-$21 54 15.90 & 16.73 & 14.95 & 13.19 &             \dots & $-$187.2$\pm$0.1 & 	     \dots &	        \dots & RGB \\
 251 & 20 05 58.88 & $-$21 56 42.30 & 16.78 & 14.97 & 13.18 &             \dots & $-$183.9$\pm$0.1 & 	     \dots &	        \dots & RGB \\
1459 & 20 06 03.19 & $-$21 56 19.60 & 16.74 & 15.19 & 13.60 &             \dots & $-$193.5$\pm$0.1 & 	     \dots &	        \dots & RGB \\
 442 & 20 06 07.18 & $-$21 55 39.80 & 16.92 & 15.30 & 13.73 &             \dots & $-$174.3$\pm$0.1 & 	     \dots &	        \dots & RGB \\
 901 & 20 06 17.33 & $-$21 52 20.90 & 16.90 & 15.20 & 13.57 &             \dots & $-$184.2$\pm$0.1 & 	     \dots &	        \dots & RGB \\
 483 & 20 06 22.55 & $-$21 55 29.30 & 16.94 & 15.33 & 13.77 &             \dots & $-$186.5$\pm$0.1 & 	     \dots &	        \dots & RGB \\
 655 & 20 06 08.80 & $-$21 54 34.90 & 16.93 & 15.33 & 13.76 &             \dots & $-$189.4$\pm$0.1 & 	     \dots &	        \dots & RGB \\
239 & 20 06 02.23 & $-$21 56 46.35 & 17.02 & 15.53 & 14.01 &  $-$201.2$\pm$1.6 & $-$194.3$\pm$0.1 & 5.34$\pm$0.08 & $-$1.01$\pm$0.12 &	RGB \\
1251 & 20 06 06.21 & $-$21 55 02.56 & 17.08 & 15.59 & 14.09 &  $-$177.2$\pm$1.8 & $-$177.1$\pm$0.1 & 5.53$\pm$0.08 & $-$0.92$\pm$0.12 &	RGB \\
 512 & 20 06 01.08 & $-$21 55 19.81 & 17.08 & 15.61 & 14.11 &  $-$191.0$\pm$0.5 & $-$195.1$\pm$0.1 & 5.37$\pm$0.10 & $-$0.97$\pm$0.12 &	RGB \\
 486 & 20 06 12.76 & $-$21 55 27.40 & 17.22 & 15.76 & 14.32 &             \dots & $-$180.6$\pm$0.1 & 	     \dots &	        \dots & RGB \\
 612 & 20 06 01.89 & $-$21 54 46.30 & 17.15 & 15.81 & 14.42 &             \dots & $-$202.6$\pm$0.1 & 	     \dots &	        \dots & RGB \\
 876 & 20 06 08.04 & $-$21 52 45.44 & 17.34 & 15.91 & 14.48 &  $-$191.9$\pm$1.5 & $-$187.0$\pm$0.1 & 5.28$\pm$0.09 & $-$0.93$\pm$0.12 &	RGB \\
1464 & 20 06 06.28 & $-$21 56 05.02 & 17.38 & 16.03 & 14.66 &  $-$181.2$\pm$1.1 & 	     \dots & 5.09$\pm$0.10 & $-$0.98$\pm$0.12 &	RGB \\
 732 & 20 06 06.40 & $-$21 54 11.12 & 17.82 & 16.50 & 15.15 &  $-$177.2$\pm$1.8 & 	     \dots & 4.84$\pm$0.19 & $-$0.96$\pm$0.14 &	RGB \\
1286 & 20 06 00.16 & $-$21 54 43.45 & 17.86 & 16.62 & 15.31 &  $-$194.7$\pm$0.7 & 	     \dots & 4.83$\pm$0.09 & $-$0.93$\pm$0.12 &	RGB \\
 887 & 20 06 06.12 & $-$21 52 36.51 & 17.91 & 16.68 & 15.39 &  $-$189.4$\pm$0.9 & 	     \dots & 4.89$\pm$0.11 & $-$0.89$\pm$0.13 &	RGB \\
 827 & 20 06 09.62 & $-$21 53 24.57 & 18.04 & 16.79 & 15.49 &  $-$176.8$\pm$1.5 & 	     \dots & 4.80$\pm$0.08 & $-$0.90$\pm$0.12 &	RGB \\
 856 & 20 06 02.76 & $-$21 53 02.00 & 18.30 & 17.16 & 15.94 &  $-$177.2$\pm$2.2 & 	     \dots & 4.30$\pm$0.39 & $-$1.01$\pm$0.20 &	RGB \\
 783 & 20 06 04.79 & $-$21 53 46.71 & 18.58 & 17.49 & 16.34 &  $-$202.6$\pm$2.4 & 	     \dots & 4.08$\pm$0.18 & $-$1.01$\pm$0.14 &	RGB \\
1115 & 20 05 58.10 & $-$21 57 34.05 & 18.76 & 17.71 & 16.53 &  $-$208.0$\pm$2.1 & 	     \dots & 4.29$\pm$0.20 & $-$0.87$\pm$0.14 &	RGB \\
 438 & 20 06 04.29 & $-$21 55 40.83 & 17.03 & 15.56 & 14.13 &  $-$186.1$\pm$0.6 & 	   \dots & 5.24$\pm$0.09 & $-$1.04$\pm$0.12 &	AGB \\
1152 & 20 06 07.84 & $-$21 56 13.84 & 17.36 & 16.15 & 14.88 &  $-$185.0$\pm$0.9 & 	   \dots & 4.82$\pm$0.09 & $-$1.06$\pm$0.12 &	AGB \\
1106 & 20 06 13.08 & $-$21 57 59.43 & 18.46 & 17.57 & 16.59 &  $-$187.3$\pm$1.1 & 	   \dots & 3.72$\pm$0.15 & $-$1.14$\pm$0.13 &	RHB \\
1328 & 20 06 09.09 & $-$21 53 57.87 & 17.93 & 17.54 & 17.06 &  $-$193.9$\pm$5.9 & 	   \dots &  \dots	   & \dots	    &	BHB \\
\hline
 \hline                  
\end{tabular}
\begin{list}{}{}
\item[$^a$] Taken from \citet{Kravtsov2007}
\end{list}
\end{table*}

The lay-out of FORS2 with its two separate CCD chips can, in principle, lead to systematic offsets in radial velocities if measured 
on different chips \citep[e.g.,][]{Fraternali2009}. In order to ascertain that our data are not affected by such systematics, we 
computed the kinematics of the subsamples from chip 1 and 2 separately and find only a marginal (0.7$\sigma$) difference 
of the mean systemic velocity as computed from either chip. We deem chip-offsets redundant for the remainder of this work and 
will further comment on this option in the sections presenting our results below.
\subsection{M\,75: Combined sample}
{ In the following,  we determined mean heliocentric radial velocities and velocity dispersions  
 in a maximum likelihood sense, by optimizing the probability $\mathcal{L}$ that the given ensemble of stars with 
 velocities $v_i$ and errors $\sigma_i$ are drawn from a population with mean velocity $<$$v$$>$ and dispersion $\sigma$   \citep[e.g.,][]{Walker2006}:
 \begin{equation}
 \mathcal{L}\,=\,\prod_{i=1}^N\,\left( \, 2\pi\,(\sigma_i^2 + \sigma^2 \, )  \right)^{-\frac{1}{2}}\,\exp \left(-\frac{(v_i \,- <v>)^2}{2\,\left(\sigma_i^2 + \sigma^2\right)} \right) 
 \end{equation}
}

From the 21 stars observed with FORS alone, we thus derive a mean heliocentric radial velocity 
of  $-188.4\pm1.8$\ km\,s$^{-1}$ and a velocity dispersion 
$\sigma$ of 8.1$\pm$1.3 km\,s$^{-1}$, both of which are consistent with earlier results \citep{Pryor1993,Harris2010}. 
Similarly, the MIKE sample of 18 stars returns a value of  $-186.2\pm1.9$ km\,s$^{-1}$ with dispersion of 7.8$\pm$1.3 km\,s$^{-1}$, in excellent agreement
with the aforementioned values. 

Seven stars have been targeted in common between both campaigns.
Here, the velocities agree excellently for three of those, while the difference is larger, up to 7 km\,s$^{-1}$ for the remainder. 
Such inconsistencies could, in principle indicate that those stars are in binaries; however, 
the temporal sampling of the observing runs is insufficient to test for more systematic variations.
Overall, the mean difference between the MIKE and FORS values  is 0.7 km\,s$^{-1}$ with a large scatter of 3.9 km\,s$^{-1}$. 
These deviations are independent on the stars' locations in the GC, their position in the CMD, and neither subset shows systematically lower or larger values alone.

To construct a final, {\em bona-fide} sample, we shifted the FORS velocities by the difference in mean velocity, thereby tying them to the more precise
MIKE scale. 
As a second test, we rejected those common stars with velocities that differ by more than the combined error bar and, 
for the remaining three overlapping stars, we computed an error-weighted mean value. 
We emphasize that the results of the kinematic analysis below are
insensitive to the in- or exclusion of the common stars, or the employed combination scheme so that we will continue with the 
 sample after shifting to a common velocity scale { and removal of the binary candidates} \citep[see also][]{Koch2007}. 
\subsection{NGC\,6426: VLT/FLAMES}
The central regions of this GC are heavily affected by 
crowding, which is unfortunate for efficient multi-object fibre spectroscopy. 
For target selection, we relied on infrared photometry from the Two Micron All Sky Survey 
 \citep[2MASS;][]{Cutri2003}. Furthermore, Hubble Space Telescope (HST) photometry is available in  V and I for a pencil beam toward this GC  \citep{Dotter2011}, which 
contains most of our target stars. The corresponding CMD is  shown in the right panel of Fig.~1. 

The observations were carried out using the HR13 grating of the GIRAFFE spectrograph as part of the 
 Fibre Large Array Multi Element Spectrograph \citep[FLAMES;][]{Pasquini2002} at the ESO/VLT.  
 This setting provides a wavelength coverage of 6100--6400 \AA~at a resolving power of 20,000, where 125 fibres were placed on stars.
 
After the loss of two entire runs over two ESO periods due to bad weather,  data could finally be accrued on the two nights of Aug. 21 and Sep. 19, 2017. 
While originally conceived as an abundance-project, only two out of the six granted  observing blocks were taken, 
which are the base for the present, kinematic study, for which the lower SNR is sufficient. 
Here, the total exposure time was 1.4 hours, split into 4$\times$1240 s exposures. 
The data were reduced using ESO standard routines. 

As above, we measured radial velocities  from a cross-correlation of the spectra 
against a synthetic template spectrum with stellar parameters representative of the target stars. 
The errors returned by  {\em fxcor} were re-scaled using the (up to four) repeat observations following the 
formalism of  \citet{Vogt1995} and \citet{Koch2007}, resulting in a median velocity error of 1.5 km\,s$^{-1}$.
The resulting measurements and target properties are listed in Table~2.
\begin{table*}[htb]
\caption{Properties of the NGC\,6426 member stars from FLAMES}              
\centering          
\begin{tabular}{cccccccccc}     
\hline\hline       
  & $\alpha$ & $\delta$ & V & V$-$I & K & J$-$K & SNR & v$_{\rm HC}$  & r   \\
\raisebox{1.5ex}[-1.5ex]{ID\tablefootmark{a}} & \multicolumn{2}{c}{(J2000.0)} & [mag] & [mag] & [mag] & [mag] & [px$^{-1}$] & [km\,s$^{-1}$] & [$\arcmin$] \\
\hline
 385$\rlap{\tablefootmark{b}}$ & 17:44:56.70 &  3:09:39.34 & 15.327 &  1.20 & 12.302 &  0.80 &  51 & $-$209.2$\pm$0.9 & 0.74\\ 
 777 & 17:44:56.81 &  3:10:47.54 & 16.067 &  1.21 & 12.991 &  0.82 &  42 & $-$205.9$\pm$0.6 & 0.78\\ 
2075 & 17:44:52.09 &  3:09:40.69 & 16.681 &  1.19 & 13.819 &  0.70 &  35 & $-$208.1$\pm$0.9 & 0.84\\ 
1198 & 17:44:52.85 &  3:09:18.07 & 16.868 &  1.11 & 14.008 &  0.68 &  30 & $-$213.2$\pm$0.9 & 1.02\\ 
 168 & 17:44:49.00 &  3:09:06.44 & 16.956 &  1.03 & 14.384 &  0.76 &  26 & $-$213.5$\pm$1.4 & 1.80\\ 
1961 & 17:44:57.21 &  3:08:56.39 & 17.125 &  1.09 & 14.413 &  0.70 &  25 & $-$212.1$\pm$1.2 & 1.41\\ 
 196 & 17:44:53.02 &  3:11:19.68 & 17.359 &  1.07 & 14.728 &  0.53 &  21 & $-$207.8$\pm$1.3 & 1.20\\ 
1898 & 17:44:56.52 &  3:11:37.31 & 17.369 &  1.08 & 14.648 &  0.80 &  20 & $-$200.1$\pm$3.0 & 1.48\\ 
 618 & 17:44:53.04 &  3:10:49.33 & 17.401 &  1.12 & 14.637 &  0.76 &  17 & $-$211.3$\pm$1.2 & 0.74\\ 
 842 & 17:44:55.54 &  3:11:09.23 & 17.417 &  1.10 & 14.519 &  0.82 &  20 & $-$208.7$\pm$2.1 & 0.97\\ 
 919 & 17:44:54.14 &  3:09:23.30 & 17.448 &  1.10 & 14.581 &  0.76 &  19 & $-$210.4$\pm$1.2 & 0.83\\ 
2169 & 17:44:48.34 &  3:09:22.05 & 17.553 &  1.02 & 15.016 &  0.82 &  19 & $-$207.6$\pm$1.5 & 1.80\\ 
 328 & 17:44:57.24 &  3:09:54.03 & 17.626 &  1.09 & 14.819 &  0.82 &  19 & $-$212.9$\pm$1.5 & 0.70\\ 
 494 & 17:45:00.17 &  3:11:22.72 & 17.716 &  1.03 & 15.098 &  0.74 &  15 & $-$203.6$\pm$2.7 & 1.80\\ 
1211 & 17:44:51.41 &  3:10:18.61 & 17.776 &  1.09 & 15.127 &  0.74 &  13 & $-$210.5$\pm$2.1 & 0.83\\ 
1575 & 17:44:51.26 &  3:11:24.78 & 17.929 &  1.04 & 15.287 &  0.70 &   9 & $-$207.5$\pm$4.6 & 1.48\\ 
 268 & 17:44:49.30 &  3:11:07.77 & 17.948 &  1.08 & 15.322 &  0.62 &  14 & $-$209.7$\pm$2.9 & 1.63\\ 
1908 & 17:44:58.00 &  3:10:46.77 & 18.065 &  1.04 & 15.526 &  0.59 &  15 & $-$209.2$\pm$1.4 & 1.00\\ 
 366 & 17:44:51.03 &  3:10:03.70 & 18.430 &  1.03 & 15.526 &  0.72 &  11 & $-$211.6$\pm$4.2 & 0.93\\ 
  89 & 17:44:43.82 &  3:07:00.26 & \ldots &\ldots & 14.903 &  0.71 &   5 & $-$214.6$\pm$6.3 & 4.20\\ 
 162 & 17:44:42.53 &  3:10:55.34 & \ldots &\ldots & 14.157 &  0.70 &  28 & $-$210.9$\pm$1.9 & 3.12\\ 
 995 & 17:44:53.75 &  3:12:56.49 & \ldots &\ldots & 15.140 &  0.77 &  15 & $-$207.8$\pm$2.2 & 2.74\\ 
1159 & 17:44:50.69 &  3:08:02.66 & \ldots &\ldots & 15.588 &  0.56 &  17 & $-$210.4$\pm$1.7 & 2.39\\ 
1529 & 17:45:04.71 &  3:09:39.16 & \ldots &\ldots & 15.065 &  0.57 &  18 & $-$211.6$\pm$1.4 & 2.56\\ 
\hline
\hline
\end{tabular}
\tablefoot{
\tablefoottext{a}{IDs based on our 2MASS input catalog.}
\tablefoottext{b}{This is star 14853 in \citet{Hanke2017}.}
}
\end{table*}

The plot of velocity vs. radial distance from the cluster center  in Fig.~3 
shows a clear signal of the GC population at a heliocentric velocity of $-209.7\pm0.5$ km\,s$^{-1}$
that is significantly detached from 
the Milky Way foreground. This value is considerably lower than that listed in the \citet{Harris2010} catalogue based on 
low-resolution spectra of a handful of stars. 
{ 
As for M\,75,  we also combined our sample with the MIKE data of \citep{Hanke2017} by shifting the FLAMES velocities to the higher-resolution MIKE scale. 
This yields} a final sample of 27 member candidates
with a dispersion of $1.9\pm0.4$ km\,s$^{-1}$, 
and we discuss their kinematic properties in the following. 
\begin{figure}
\centering
\includegraphics[width=1\hsize]{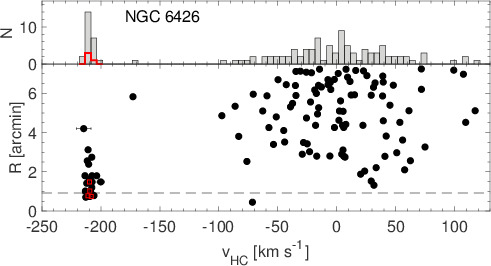}
\caption{Heliocentric radial velocity against radial distance (bottom panel)  for the entire NGC\,6426 observations. The histogram in the top panel indicates the peak of the GC at $-$210 km\,s$^{-1}$.
The dashed line in the bottom panel  shows the cluster's half-light radius. Red symbols and the red histogram refer to the MIKE sample of \citet{Hanke2017}.}
\end{figure}
\section{M\,75: Calcium Triplet metallicities}
As a more detailed chemical abundance analysis of this GC exists from the MIKE spectra of \citet{Kacharov2013}, 
here, we determined the equivalent widths (EWs) of the  CaT lines  from the FORS2 spectra 
as another means of  membership assessment. 
To this end, we fitted Gaussian plus Lorentz line profiles \citep{Cole2004}  to each line using line and continuum band passes from \citet{Armandroff1988}. 
Adopting the calibrations of line strength onto metallicity by \citet{Rutledge1997a,Rutledge1997b}, we find that 18 of the 21 likely radial velocity members also cluster in 
abundance space (Fig.~2, middle panel) around [Fe/H]$_{\rm CaT} = -1.1$ dex on the scale \citet{Carretta2009Fe}. This compares favourably with previous 
photometric estimates of $-0.7$ to $-1.3$ 
dex \citep{Catelan2002,Kravtsov2007,Kacharov2013}. 
The other three stars with lower metallicities at M\,75's systemic velocity can still be considered likely members: these are the reddest bright objects at V$-$I$>$1.9 mag. 
Given the inflection of this metal-rich GC's RGB near the tip these are likely luminous AGB stars with 
notably narrow (CaT) absorption lines, for which the linear CaT calibration becomes unreliable \citep[e.g.,][]{Garnavich1994}. 
Moreover, the data of \citet{Corwin2003} indicate that, at least two out these three, are variable stars.

As a result, the mean metallicity of $-1.00\pm0.04$ dex with an intrinsic dispersion of 0.07$\pm$0.05 dex is slightly more metal-rich than the 
high-resolution, high-SNR MIKE study of \citet{Kacharov2013}, who found a mean [Fe\,{\sc i}/H] of $-1.16\pm0.02$.
This systematic over-estimate from the low-resolution FORS2 spectra is also seen in the one-to-one comparison of the stars that overlap between the both samples. 
This discrepancy is considerably alleviated when comparing our results to the metallicity scale from ionized iron lines: here, \citet{Kacharov2013} find 
a GC mean of  [Fe\,{\sc ii}/H] = $-0.98\pm0.03$ dex (random) $\pm$0.16 dex (systematic) with a 1$\sigma$ spread of 0.13 dex, which is fully compatible with our 
present finding. This strongly attests to favouring the use of Fe\,{\sc ii} as a prime metallicity scale \citep[e.g.,][]{KraftIvans2003,Hanke2017}, as 
it is also less prone to NLTE-effects. 
\section{General kinematics}
Based on our measurement of 32 (27)  heliocentric velocities in M\,75 and NGC\,6426, respectively, we 
determined the global kinematic properties of the GCs (as listed in Table~3) from the {\em bona fide}, combined samples.
\begin{table}[htb]
\caption{Global and kinematic properties of the two targeted GCs}              
\centering          
\begin{tabular}{cccc}     
\hline\hline       
 & \multicolumn{2}{c}{Value} &  \\
 \cline{2-3}
\rb{Parameter} & M\,75 & NGC\,6426 & \rb{Reference\tablefootmark{b} } \\
\hline
$l$  [$\degr$] & \phs20.30 & 28.09 & 1 \\[0.5ex]
$b$  [$\degr$] & $-$25.75 & 16.23 & 1 \\[0.5ex]
R$_{\rm GC}$ [kpc] & 14.7 & 14.4 & 1 \\[0.5ex]
R$_{\odot}$ [kpc] & 20.9 & 20.6 & 1 \\[0.5ex]
$r_h$ [$\arcmin$] & 0.46 & 0.92&1 \\[0.5ex]
$r_t$ [$\arcmin$] &  5.7 & 13.0 & 1\\[0.5ex]
$[$Fe/H$]$ & $-1.13$/$-$1.16 & $-2.34$ & 2,3,4 \\[0.5ex]
N\tablefootmark{a} & 32 & 27 & 2,4\\[0.5ex]
$v_{\rm HC}$ [km\,s$^{-1}$] & $-$186.2$\pm$1.5 & $-$209.7$\pm$0.5 & 2 \\[0.5ex]
$\sigma$ [km\,s$^{-1}$] & 8.2$\pm$1.1 & 1.9$\pm$0.4 & 2 \\[0.5ex]
$A$ [km\,s$^{-1}$] & 5.0$\pm$0.9 & 3.9$\pm$1.8 & 2 \\[0.5ex]
PA [$\degr$] & $-$15$\pm$30 &  $281^{+36}_{-25}$ &  2 \\[0.5ex]
A$_{\rm rot}$ [km\,s$^{-1}$]  & 4.5$\pm$2.1 &  $2.0^{+1.1}_{-0.9}$ & 2; Eq.~2 \\[0.5ex]
$R_{\rm peak}$ [$\arcmin$] & $1.6^{+1.6}_{-1.1}$ &  $2.1^{+1.5}_{-1.2}$ & 2; Eq.~2 \\[0.5ex]
$\sigma_0$  [km\,s$^{-1}$] & $14.1^{+2.4}_{-2.0}$ & $2.4^{+0.7}_{-0.5}$ & 2; Eq.~3 \\[0.5ex]
{$\mu_{\alpha\,\cos\,\delta}$ [mas\,yr$^{-1}$]} & $-0.39^{+0.79}_{-0.63}$ & $-1.82^{+0.31}_{-0.34}$ & 2,5 \\[0.5ex]
{$\mu_{\delta}$ [mas\,yr$^{-1}$]} & $-2.76^{+0.52}_{-0.60}$ &  $-2.98^{+0.34}_{-0.38}$ & 2,5 \\
\hline
\hline
\end{tabular}
\tablefoot{
\tablefoottext{a}{Number of member stars in the respective analysis}
\tablefoottext{b}{References: (1): \citet{Harris2010}; (2): This work; (3): \citet{Kacharov2013}; (4): \citet{Hanke2017}; (5): \citet{GaiaDR2}.
}
}
\end{table}

For M\,75, we determined a mean heliocentric velocity of 
$-186.2\pm1.5$ km\,s$^{-1}$ with a velocity dispersion of 8.2$\pm$1.1 km\,s$^{-1}$.
While this dispersion seems high for a GC at a first glance, it has to be kept  in mind that M\,75 is a massive system  so that our measured value is to be expected 
and fully compatible with the high values for $\sigma$ found in other luminous,  dense systems \citep[Fig.~4; ][]{Pryor1993}.  
\begin{figure}
\centering
\includegraphics[width=0.9\hsize]{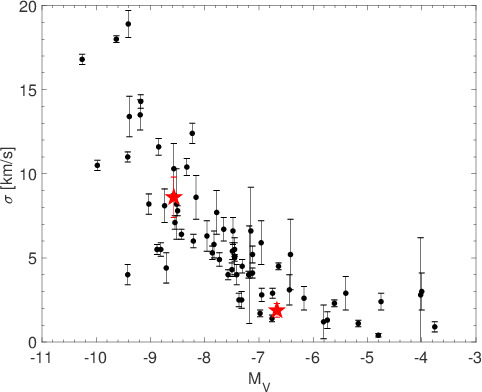}
\caption{Global velocity dispersion vs. absolute magnitude, using data from \citet{Pryor1993} and \citet{Harris2010}, 
and with the latest measurements of low-mass GCs
from \citet{Koch2017ESO,Koch2017FSR,Koch2018Gaia}. Targets of this study are indicated 
by red stars.}
\end{figure}

For NGC\,6426, we find a mean  velocity and dispersion of $-209.7\pm0.4$ km\,s$^{-1}$ and 1.9$\pm$0.4 km\,s$^{-1}$, respectively.
The low, latter value is in line with this GC's lower mass, as also suggested by Fig.~4. 
\citet{Hanke2017} found a mean systemic velocity that is lower by 2.4$\pm$0.8 km\,s$^{-1}$ that was, however, based on four stars.  Accordingly, their velocity dispersion 
is considerably lower than our value, at 1.0$\pm$0.4 km\,s$^{-1}$. 
One star in the present sample is in  common with \citet{Hanke2017}. Here, the velocity from their MIKE spectra differs by 1.7 km\,s$^{-1}$ (1.2$\sigma$). 
Since the latter work was focused on a high-resolution chemical abundance analysis, this discrepancy may hint at an underestimate of the 
velocity errors, which were not crucial for the chemical purpose. 
Likewise, the different result by \citet{Dias2015} of a mean velocity of 
$-242\pm11$ km\,s$^{-1}$ can be understood in terms of the small sample size (at five stars) and a low spectral resolution of R$\sim$2000.

Since Galactic foreground stars at similar velocities can, in principle, inflate  velocity dispersion measurements and corrupt kinematic 
analyses \citep[e.g.,][]{Walker2009}, we estimated the amount of contamination via the 
Besan\c{c}on Galaxy model \citep{Robin2003}. As a result we do not expect a single foreground interloper 
within the CMD selection criteria and at velocities below $-100$ km\,s$^{-1}$ drawn from samples like our NGC\,6426 observations. The same holds for M\,75. 
\section{Rotation analysis}
As a first step, we followed the standard procedure of dividing the data into bins of position angle and computing the mean velocity of the respective subsamples
\citep[e.g.,][]{Koch2007,Kacharov2014}. 
The resulting difference in mean velocity  from either side of a line passing through the GC center at that angle 
then yields the projected rotation curves in Fig.~5. 
  These can be described by a simplistic sinusoidal 
with amplitude $A$ and position angle (PA, counting North through East) as $\Delta$v\,=\,$A \sin\left(\varphi+{\rm PA}\right)$, where $\varphi$ is the
position angle of each star with respect to the cluster center. 
As the spatial distribution of the stars  in the bottom panel of Fig.~5 indicates, the main source of uncertainty in this analysis will be the 
small-number statistics of the samples. 
\begin{figure}
\centering
\includegraphics[width=1\hsize]{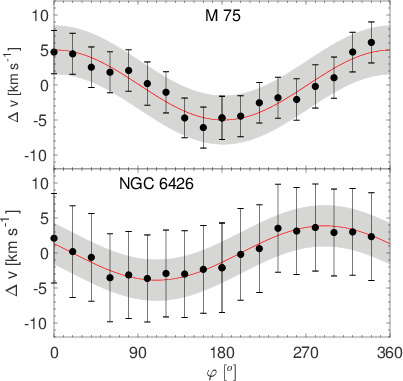}\\
\vspace{2ex}
\includegraphics[width=1\hsize]{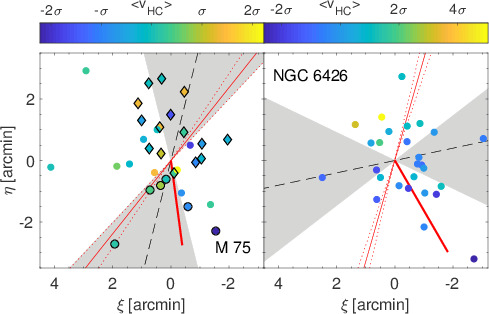}
\caption{Top panels: Rotation curves and best-fit sinusoidal for both GCs. The bottom panels
show the location of the target stars, colour-coded by radial velocity in units of their dispersion. 
Here, diamonds refer to M\,75 targets on chip \#1 on FORS2, circles are stars on chip \#2, and MIKE targets are shown without extra delimiters. 
The gray shaded areas indicate the error margin of the rotation axis, which itself is shown as a dashed line.
Each  clusters'  major axis position angles from \citet{ChenChen2010} are plotted as thin red lines; { 
finally, the direction of the Gaia DR2 proper motion is shown via a thick red line.}}
\end{figure}

Next we performed a more advanced Bayesian fit to the discrete kinematic data of the two GCs  \citep[][see also \citealt{Kacharov2014}]{Mackey2013}, 
accounting simultaneously for both the rotation profile 
$V_{\rm rot}(X_{\rm PA}$) (Eq. 2) and velocity dispersion profile $\sigma(R)$ (Eq. 3). The former parameterization adopts a cylindrical rotation and violent relaxation. 
Overall, this approach and the associated likelihood distributions follows closely the method described by \citet[][their Eqs.~2,5]{Cordero2017}. 
\begin{eqnarray}
V_{\rm rot}\,\sin i\,&=&\,\frac{2\,A_{\rm rot}}{R_{\rm peak}}\,\frac{X_{\rm PA}}{1\,+\,\left(X_{\rm PA}/R_{\rm peak}\right)^2} \\
\sigma(R)\,&=&\,\sigma_0\,\left(\,1\,+\,(R/a)^2\,\right)^{-1/4} 
\end{eqnarray}
The adopted dispersion model follows the prediction for a Plummer sphere with one free parameter -- the central velocity dispersion $\sigma_0$. 
We kept the half-mass radius $a$ fixed at the half-light radius divided by a factor of 1.3.  

In contrast, the rotation parameterization  has three free parameters: the position angle (PA) of the rotation axis, the amplitude of rotation ($A_{\rm rot}$), 
and the distance of the peak rotation from the cluster's centre, $X_{\rm rot}$. 
As per Eq.~2, this rotation curve is given as a function of the distance along the major axis, also accounting for 
the unknown inclination $i$.  
As we found that the position of the peak rotation is not well constrained by the data, we used a normal distribution with a mean at the cluster's centre and a standard deviation of 2$\arcmin$ as a prior on this parameter.
In practice, we sampled the joint posterior distribution of the four free parameters 
using the affine-invariant MCMC algorithm \citep{Goodman2010,Foreman-Mackey2013}. 
The resulting posteriori probability distributions are shown in Fig.~6, and in the following Section we summarize the resulting 
free parameters in terms of their median values and  adopt the 15.9\% and 84.1\% percentiles as lower and upper error bounds.
{ While the median values indicate a marginal level of rotation in either object, the full probability distributions suggest that an absence of rotation 
cannot be fully excluded.}
\begin{figure}
\centering
\includegraphics[width=1\hsize]{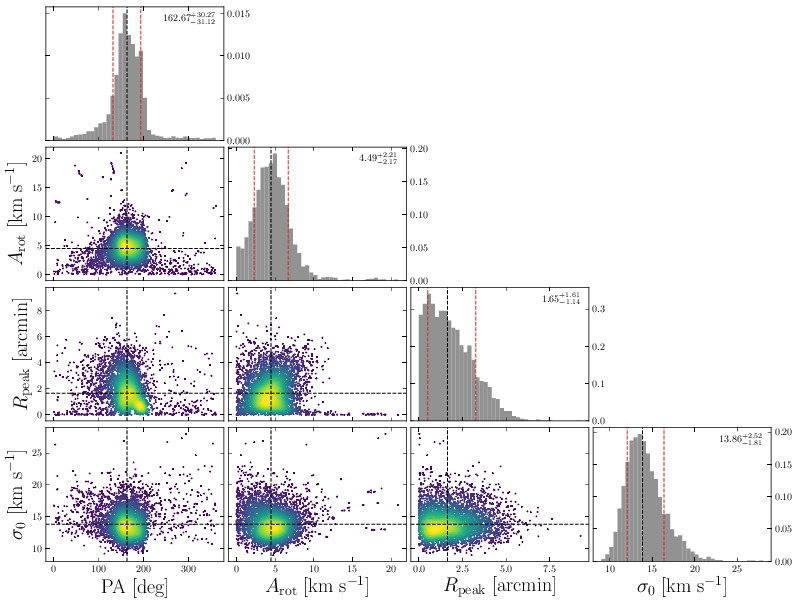}
\includegraphics[width=1\hsize]{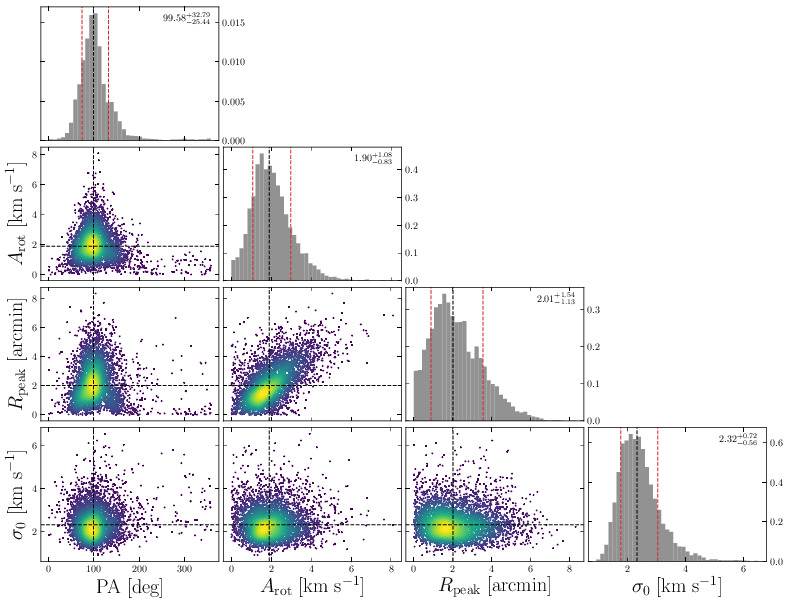}
\caption{Posterior probability distributions of our rotational analysis, with parameters as defined in Eqs.~1,2. The top and bottom panels show the results for 
M\,75 and NGC\,6426, respectively.}
\end{figure}
\subsection{M\,75}
Our tests for rotation suggest that M\,75's rotation axis is inclined at a position angle of $-15\degr\pm30\degr$. 
Using 2MASS photometry and density distributions, \citet{ChenChen2010} determined the morphological parameters for almost the entire Galaxy's GC population. 
For M\,75, their study implied a major axis position angle\footnote{We note that \citet{ChenChen2010} reported their parameters in Galactic coordinates, leading to an original angle with respect to the North 
Galactic Pole of 148$\degr$.} of $40\pm5\degr$. 
Since morphological distortions would be manifest in a flattening that is aligned with the minor axis, 
we will consider in the following rather the minor axis position angle, which for M\,75 is thus  $-50\degr$. 
In this regard, our kinematic rotation axis is offset to the cluster's minor axis by 35$\degr$, but it should be kept in mind that 
the actual flattening of M\,75 is very small, at an axis ratio of 0.92$\pm$0.03 \citep{ChenChen2010}, and that our kinematic analysis is aggravated by 
small-number statistics. However, as the distinction in Fig.~5 (bottom left) of stars on the two FORS2 chips indicates, the 
preferential rotation direction is not aligned with the instrument so that chip-offsets are unlikely to be the cause of the rotation signal. 
This is bolstered by the independent confirmation of the additional MIKE data.

The amplitude of the sinusoidal in Fig.~4, which is statistically on the order of twice the overall average rotation signal \citep{Bellazzini2012}, 
is $A=5.0\pm0.9$ km\,s$^{-1}$, while the more detailed Bayesian  
fit of Eq.~2 to the non-binned profile yielded a half-amplitude of $A_{\rm rot}=4.5\pm$2.1 km\,s$^{-1}$ around an axis that is offset from the center by some 
1.6$\arcmin$ (Fig.~7; Table~3). 
At the GC's half-light radius of 0.5$\arcmin$ \citep{Harris2010}, this would correspond to  three such radii and would be remarkably large. 
{ Considering the poor sampling of our data in the central regions of M\,75, we do not interpret this value any further.} 	
\begin{figure}
\centering
\includegraphics[width=1\hsize]{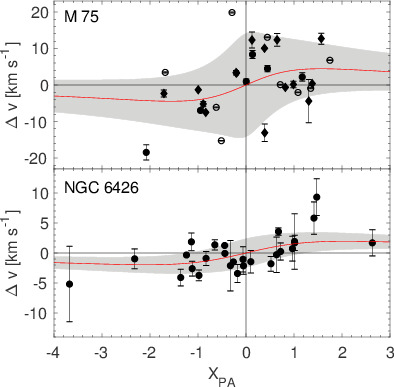}
\caption{Rotation curve and best-fit curve \`a la Eq.~2. The gray-shaded area indicates the margins imposed by the radial velocity dispersion profile (Eq.~3).
Different symbols for M\,75 reflect targets on different FORS2 chips (filled) and those observed with MIKE (open).}
\end{figure}
\subsection{NGC\,6426}
The best-fit position angle for this GC is $281\degr$. This compares to a {\em minor} axis position angle from the morphological analysis of 
\citet{ChenChen2010}
of $253\degr\pm3\degr$, which is marginally consistent with our measurement. 
On the other hand, our derived amplitudes of $A=3.9\pm1.8$ and $A_{\rm rot}=2.0^{+1.1}_{-0.9}$  km\,s$^{-1}$, the latter with a
nominal peak at $2.1\arcmin$$^{+1.5}_{-1.2}$ (thus well outside the cluster core, at $r_c=0.26\arcmin$), { are clearly a limited representation of NGC\,6426's 
kinematics.} 
\section{Orbits}
{ In order to  obtain the full kinematic information on our clusters, we 
derived their proper motions using the recent second data release, DR2, of Gaia  \citep{GaiaDR2}.
To identify the {\em bona fide} cluster signature in proper motion space, we selected stars within 1.5 times $r_h$, vetting those with significant parallaxes as
near-by foreground stars, and we demanded that 
the errors in either direction are $\sigma_{\mu}<0.5$ mas\,yr$^{-1}$. This way, we 
utilized 161 and 103 objects towards M\,75 and NGC\,6426, respectively.
The resulting median values (with 15.9\% and 84.1\% percentile uncertainties) we find are ($\mu_{\alpha\,\cos\,\delta},\mu_{\delta}$)=($-0.4^{+0.8}_{-0.6},-2.8^{+0.5}_{-0.6}) $ mas\,yr$^{-1}$ for M\,75 and 
($-1.8\pm0.3,-3.0^{+0.3}_{-0.4}$) mas\,yr$^{-1}$ for NGC\,6426.
These values are in excellent agreement with the recent HST study of \citet{Sohn2018} for NGC 6426 and with the values for M\,75 of \citet{Chemel2018}, 
in turn based on a variety of surveys including Gaia DR1. 
It is interesting to note that, for M\,75, the direction of motion is consistent with the rotation axis, while for NGC\,6426 is it almost perpendicular.
Our derived proper motions were then fed into the Galactic potential of \citet{Dehnen1998}, which contains contributions from the halo and disk, and 
spherical bulge.
The orbital integrations were carried out backwards for 12 Gyr and the resulting projections are shown in Fig.~8.
\begin{figure}
\centering
\includegraphics[width=1\hsize]{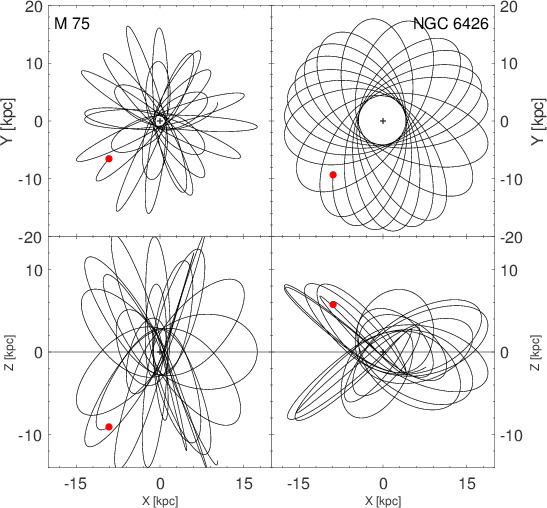}
\caption{Orbit projections as a result of a backward computation. Here, only a timeframe of 6 Gyr is depicted. The present location of the GCs is shown as red 
point. The cross and line denote the Galactic center and plane.}
\end{figure}

Both GCs show typical motions with apocenters of 17.5 kpc (M\,75) and 19.4 kpc (NGC\,6426) that lie close to their current Galactocentric distances. 
At 0.4 Gyr each, their orbital periods are fairly short. 
The eccentricities of these outer halo objects are large, at $e=0.63$ for NGC\,6426, and even more notably for M\,75, which shows an eccentric orbit with $e$=0.87. 
The latter could indicate that M\,75 has once been accreted, as also bolstered by its younger age compared to other GCs at similar metallicities \citep{Catelan2002}. However, 
while this object lies close in projection to the disrupted Sagittarius Stream, no tidal debris around M\,75 has been observed \citep{Carballo-Bello2014}, 
and \citet{Majewski2004} argue that this GC is unlikely to be associated with Sagittarius based on its large approaching radial velocity. 
As for NGC\,6426, 
\citet{Forbes2010} note that this GC lies close to the purported orbit of the Canis Major overdensity, but given our present result, we cannot unambiguously 
constrain an in- or ex-situ origin of this metal-poor object. 
Globally speaking, the full, three-dimensional kinematics of stellar systems in the outer halo has further important implications for constraining the mass of the Milky Way \citep{Watkins2018,Helmi2018}.
} 
\section{Discussion}
The ratio of  a system's rotation amplitude and central velocity dispersion,  $A_{\rm rot}/\sigma_0$, versus its ellipticity (Fig.~9) is a representative measure for the importance of its dynamic-morphological
interaction \citep[e.g.,][]{Davies1983}.
While we were able to compute a meaningful global velocity dispersion for our two target GCs, 
{ the overall small number of sample stars aggravated a 
precise derivation of a dispersion profile}, 
and the measurement of the central velocity dispersion $\sigma_0$ in our probabilistic fitting 
should be taken with caution. 
However,  a radial dispersion gradient as seen in most GCs, e.g., when following the common Plummer profile 
\citep{Mackey2013,Cordero2017}, can be assumed. 
Moreover, many GCs are well described by a truncated \citet{King1966} profile, which is essentially flat within a few core radii. 
Thus we can state the measured, global velocity dispersion as a lower limit of $\sigma_0$, as is indeed realized in our rotational analysis.  
\begin{figure}
\centering
\includegraphics[width=1\hsize]{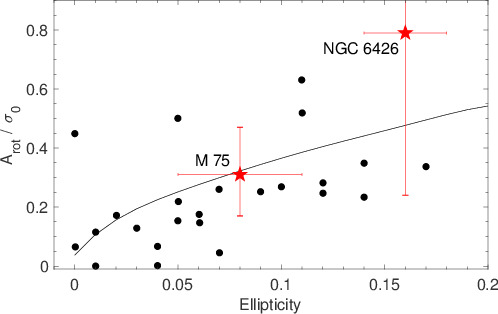}
\caption{Ratio of rotation amplitude and central velocity dispersion, according to \citet{Kacharov2014} with additional measurements from 
\citet{Bellazzini2012}. The solid line is an isotropic rotating sphere as described by \citet{Binney2005}.}
\end{figure}

{ Despite the, overall, large error bars} the outer halo GC M\,75 can be  firmly considered as a slow rotator.  
It is worth noting that M\,75 is very similar to NGC 4372 in terms of  its $A_{\rm rot}/\sigma_0$ ratio and its ellipticity, 
although the latter is less massive and more metal poor \citep{Kacharov2014,SanRoman2015}.
In the parameter space of rotation vs. metallicity, M\,75 falls fully in line with the majority of other Milky Way GCs, which is also consistent with its peculiar HB morphology.
In turn, the very metal-poor NGC\,6426 has a nominal, very large  $A_{\rm rot}/\sigma_0$ ratio of $0.8\pm0.4$, { although its large 
uncertainty renders it equally compatible with other GCs with slower rotation properties} at comparable ellipticity. 
This cluster appears as an outlier, showing a more ordered dynamics given its low metallicity, which is, however, also seen in other metal-poor GCs below $-$2 dex. 
Here it is interesting to note that the minor axis of its, relatively large, flattening is consistent  with the rotation axis 
found in the present study to within the uncertainties.

At their large distances in the halo, external tides from the Galactic disks are unlikely to play a significant dynamic role in shaping the outer halo clusters, 
and the observed, slow rotation favours internal dynamic processes as a cause for the mild observed flattening over tidal effects. 
{ Here, N-body simulations \citep[e.g.,][]{Tiongco2017} show that, as GCs dynamically evolve in a tidal field, they become progressively 
dominated by random motions while losing angular momentum. Nonetheless, even after many relaxation times and accounting for mass loss from the 
GCs, they can still be characterized by non-negligible $A_{\rm rot}/\sigma_0$ ratios.}

Moreover, \citet{Kacharov2014} conjecture that most of the slow rotators
are located 
on the  younger, presumably accreted branch in the age-metallicity space \citep{Marin-Franch2009}. 
Coupled with other evidence such as its very eccentric orbit, its younger age compared to other GCs at the same metallicity,
a large enrichment in the $r$-process elements \citep{Kacharov2013}, 
and the slow rotation pattern, 
it is likely that the more metal-rich of our targets, M\,75, is a prime example of an accreted outer halo object, although the host to its accretion has still to be
determined. 
\begin{acknowledgements}
The authors are grateful to Eline Tolstoy for helpful comments on the FORS2 data set. The anonymous referee is thanked for a constructive and helpful report . 
This work was supported  by Sonderforschungsbereich SFB 881 "The Milky Way System" (subproject A08) 
of the German Research Foundation (DFG).
\end{acknowledgements}
\bibliographystyle{aa} 
\bibliography{ms} 
\end{document}